\UseRawInputEncoding
\documentclass[12pt,letterpaper]{article}
\usepackage{amsmath,amssymb,pgf,pgfarrows,pgfnodes,float,appendix, hyperref}
\usepackage{graphicx}
\usepackage{subfigure}
\usepackage[margin=0.9in]{geometry}

\newcommand{\be}{\begin{equation}}
\newcommand{\ee}{\end{equation}}
\newcommand{\bea}{\begin{eqnarray}}
\newcommand{\eea}{\end{eqnarray}}

\title{{\rm\footnotesize \qquad \qquad \qquad \qquad \qquad \ \qquad \qquad \qquad \ \ \ \ \ \                 RUNHETC-2026-3}\vskip.5in  What is a Gravitational Path Integral? {\it or} Gravitational Path Integrals as Fluctuating Gravito-Hydrodynamics }
\author{Tom Banks\\
Department of Physics and NHETC\\
Rutgers University, Piscataway, NJ 08854\\
E-mail: \href{mailto:tibanks@ucsc.edu}{tibanks@ucsc.edu}
\\
\\
\\
\\
}
\date{}
\begin{document}
\maketitle

\begin{abstract} We show how Gravitational Path Integral formulae for various quantities that have been computed in the literature, follow from a few coarse grained hydrodynamic assumptions about the relations between space-time geometry, entropy, and fluctuations of the modular Hamiltonian of causal diamonds.  These remarks have implications for the way we think about such path integrals in relation to a more fundamental model of quantum gravity, and to questions about which space-time topologies are actually summed over in real models.  
 \end{abstract}
\maketitle

\section{Gravitation, Hydrodynamics and Path Integrals}

An enormous amount of recent work, following the seminal paper of Saad, Shenker and Stanford\cite{sss} has been devoted to deriving coarse grained quantum properties of quantum gravitational systems from Euclidean gravitational path integrals.  There has, in the present author's opinion, been a lot of unnecessary confusion about the meaning of the gravitational path integral in much of this literature.  The note\cite{tbwormhole} was an attempt to clear this up, but it has not had much impact.  The purpose of the present note is to try to make those arguments more precise.

Path integrals were invented almost simultaneously to solve two very different kinds of physical processes: the Schrodinger equation\cite{Feynman} and the time evolution of a Markov process\cite{KacWiener}.  They're relevant because in each of these situations some quantity characterizing the probability of getting from state A to state B in time t, is the sum of the same quantity for each of the possible histories that goes from A to B in time t.  In quantum mechanics that quantity is the probability amplitude, while for Markov processes it is the probability itself.  In many cases, the resulting equations resemble each other after continuing time to imaginary time.   The confusion arises because of the remarkable fact, first fully appreciated by Schwinger, that for many quantum systems one can analytically continue time to imaginary values and get expressions that are much easier to handle mathematically, and then get the correct quantum answers by analytically continuing back.

So the question arises: ``When we do Euclidean path integrals for gravity, are we doing analytically continued quantum mechanics, or are we doing statistical mechanics/hydrodynamics?".  Much of the literature, following both old ideas from the pre string theory era, and the fact that perturbative string theory and AdS/CFT treat gravitons as quantum mechanical particles, assumes the former, although it is clear that all of the non-perturbative results that have been obtained calculate coarse grained statistical quantities.  In addition, many of them violate the factorization properties expected from quantum mechanics, but not from hydrodynamic time averaged quantities.  

An entirely different point of view emerges if one follows Jacobson's seminal paper\cite{ted95}.  This paper shows that Einstein's equations (dotted into an arbitrary null vector in a Lorentzian space-time satisfying them) are the hydrodynamic equations of the entropy law 
\begin{equation} \langle K \rangle = \frac{A_{\diamond}}{4G_N} , \end{equation} for every causal diamond in the space-time.  Here $K$ is the modular Hamiltonian of the subsystem represented by the diamond, of whatever model of quantum gravity describes the global space-time.  $A$ is the maximal $d-2$ volume of space-like surfaces in a null foliation of the diamond boundary.  Implicit in this definition is an assumption that the geometry defines a particular state, which we will call the {\it empty diamond state} of that model, and that finite area diamonds define subalgebras of the operator algebra, which admit density matrices.  If general relativity is hydrodynamics, then it seems reasonable to interpret Euclidean gravitational path integrals as part of fluctuating hydrodynamics, rather than an approximate form of  analytically continued quantum gravity.

Note that this in no way contradicts the perturbative use of the Einstein-Hilbert Lagrangian to compute Feynman diagrams.  In many condensed matter systems the low energy excitations are phonons obtained by quantizing the equations of linearized hydrodynamics and doing perturbation theory around them.  In the rest of this note we will rephrase Jacobson's results in an intriguing new way and try to use the work of\cite{carlip}\cite{solo}\cite{BZ} to write down the fluctuating version of gravito-hydrodynamics.  Finally we'll try to relate this to path integrals for gravity and discuss what kind of topologies one is supposed to sum over.

\section{Jacobson Revisited: The Covariant Entropy Principle}

We have stated Jacobson's result in language he did not originally use, language that's perhaps closer to his later paper\cite{tedentangle}.  From now on we'll call the ansatz that a causal diamond has $  \langle K \rangle = \frac{A_{\diamond}}{4G_N}$ the {\it Covariant Entropy Principle} (CEP).  It's a strong form of the covariant entropy bound of\cite{fsb}.  What we know in familiar asymptotic backgrounds is that the bound is saturated when the diamond is entirely filled with a black hole.  The CEP says that this entropy is already there when the diamond is empty.  We cannot emphasize too strongly that this is the entropy of the diamond subsystem in the state of the system where the diamond is empty.

This seems puzzling until we think about it in quantum field theory (QFT) where an empty diamond means a diamond in the QFT vacuum state.  It's been known since the 1960s that the operator algebra of such a diamond is Type $III_1$ in the Murray von Neumann classification, because of the infinite entanglement caused by divergent light cone commutators.  Sorkin was the first person\cite{sorkin} to show this led to an infinite entanglement entropy proportional to the area, but it was only in the 1990s that this was associated with the Bekenstein-Hawking entropy of black holes\cite{srednickietal}.  That insight was what finally led to Jacobson's argument.  We'll see that if we follow through on this line of thought, it leads to {\it an interpretation of energy and momentum in the bulk of a diamond as an entropy deficit} - a constraint on fundamental holographic q-bits that live on the diamond boundary.  This interpretation holds locally, independent of the values of the cosmological constant (c.c.).  It has to be modified on scales larger than the radius of curvature defined by a negative c.c. because in that case, the global empty diamond state is pure.  The tensor network/error correcting code construction of the CFT ground state, shows us how this pure state can be built out of units where energy is an entropy deficit.  

Let us now turn to a rederivation of Jacobson's result from this slightly novel point of view.  We consider a point in a Lorentzian space-time and realize that it could be lying on the maximal area surfaces of many different causal diamonds.  Choose one, and vary it slightly by changing the time-like geodesic defining it, or the interval along that geodesic.  We can use the Raychaudhuri equation to compute the variation in area of the diamond.  Since we are sitting at a maximum of the area, the higher order terms in the equation are negligible and only the Ricci curvature contributes.  The variation is
\begin{equation} \delta S = R_{mn} k^m k^n / 4 G_N , \end{equation} where $k^m$ is the null direction in which the surface has moved.  Since this is happening everywhere in space-time, we make the standard assumption of hydrodynamics that it can be interpreted as coming from the flow of a conserved entropy current
\begin{equation} S_{mn} = (R_{mn} - \frac{1}{2} g_{mn} R)/4G_N , \end{equation} flowing through the surface in the direction $k^m k^n$.  We note that a possible contribution to the current from a term $constant \times g_{mn}$ gives no change in entropy.  This is because the role of the cosmological constant is to control the relation between the {\it asymptotic} limits of the two geometric invariants, maximal proper time and maximal area, of a causal diamond.  It has nothing to do with local flows of entropy.  

For the empty diamond state of a given geometry satisfying the vacuum Einstein equations, the entropy flow is zero by definition.   If we take that state as defining our system, and solutions with non-zero stress tensors (with appropriate boundary conditions) as describing the hydrodynamics of excited states of the system, then the combination of the CEP and the covariant entropy bound\cite{fsb}, imply that {\it a non-vanishing stress tensor in the bulk of a diamond is equivalent to an entropy deficit}.  This principle was first guessed at by the author and W. Fischler based on the Schwarzschild-de Sitter black hole metric, and a quantum mechanical model incorporating it was constructed by B. Fiol\cite{bfm}.  It has become the guiding principle for the construction of Holographic Space-time (HST) models\cite{hst} in asymptotically flat and asymptotically de Sitter space-times.  It should be emphasized that this is only the leading correction to the entropy of the empty diamond state when a localized excitation is introduced into a diamond.  As shown clearly by the Schwarzschild-deSitter entropy formula, there is also a positive contribution, corresponding to what we normally think of as the entropy of the localized excitation.  In many approaches to ``semi-classical" gravity, one takes into account just the area term and the localized entropy, often imagining that the latter can be computed in QFT.  For non-negative c.c. or diamonds smaller than the AdS radius for negative c.c., this ``approximation" leaves out an important term.  

The view of localized energy as an entropy deficit enables one to understand how the horizons of black holes and empty diamonds can look locally identical, yet behave so differently gravitationally.  To distinguish the two one has to look at their properties as states of a system on a larger diamond.  The empty diamond subsystem of the larger diamond will be in full equilibrium with all of the large diamond degrees of freedom, while the black hole will have a large number of large diamond q-bits frozen, enabling it to evolve as an independent system.  The gravitational attraction between the black hole and some other localized object in the large diamond is the first stage in their attempt to equilibrate with the vastly larger number of q-bits in the large diamond. It comes from virtually exciting and de-exciting frozen q-bits, which mediate interactions between the two localized objects and the majority of the large diamond degrees of freedom. Details of models that can reproduce the scaling laws of black hole physics can be found in\cite{hst}.  

Phenomenologically, hydrodynamics always requires a stochastic ``stirring force" to fit data.  Over the years, quantum theorists have realized that at least part of the origin of these statistical uncertainties in the hydrodynamic variables comes from the quantum nature of the microscopic systems that give rise to hydrodynamic flows.  In 1998, Carlip\cite{carlip} and Solodukhin\cite{solo} independently proposed a quantum model of the fluctuations on black hole horizons: a $1 + 1$ dimensional conformal field theory (CFT) on the stretched horizon (a Planck sized interval just inside the horizon), with central charge proportional to the area in Planck units.  The authors of\cite{BZ} generalized the arguments of Carlip and Solodukhin to the vicinity of the maximal area surface of any causal diamond, and argued that because the central charge was large, a cutoff CFT was adequate, as long as the cutoff on Virasoro eigenvalues was above the point where Cardy's formula was valid.  This leads to fluctuation formulae
\begin{equation} \langle (K - \langle K \rangle)^2 \rangle = \langle K \rangle  = \frac{A}{4G_N} , \end{equation}
\begin{equation} \langle S_{vv} (z) S_{vv} (w) \rangle = \frac{A}{16 G_N (z - w)^4} . \end{equation} 
 where $A$ is the $d-2$ volume of the maximal volume surface on the diamond boundary.  $S_{vv}$ is the fluctuating entropy density on the stretched horizon, and $z,w$ two points at different null coordinates near the maximal volume surface.  The first equality is satisfied if $K = L_0$, the Virasoro generator of a CFT on the stretched horizon, with a cutoff in $L_0$ eigenvalue above the stationary point that dominates the spectral sums for large central charge.  The cutoff also regulates the apparent singularity at coinciding dimensionless coordinates $z = w$.  In\cite{hilbertbundles} the cutoff on the CFT was associated with the size of causal diamond for which the hydrodynamic arguments of Jacobson, Carlip and Solodukhin break down.  This will surely be model dependent.  
 
 We are going to take these hydrodynamic properties of diamonds as basic postulates. In doing so, we're {\it postulating} the results of all Euclidean path integral computations of entropy in terms of area.  We feel that Jacobson's remarkable result is sufficient justification for this.  A simple assumption about the relationship between coarse grained quantum mechanics and geometry leads to Einstein's equations as hydrodynamics.  

\begin{figure}[h]
\begin{center}
\includegraphics[scale=0.5]{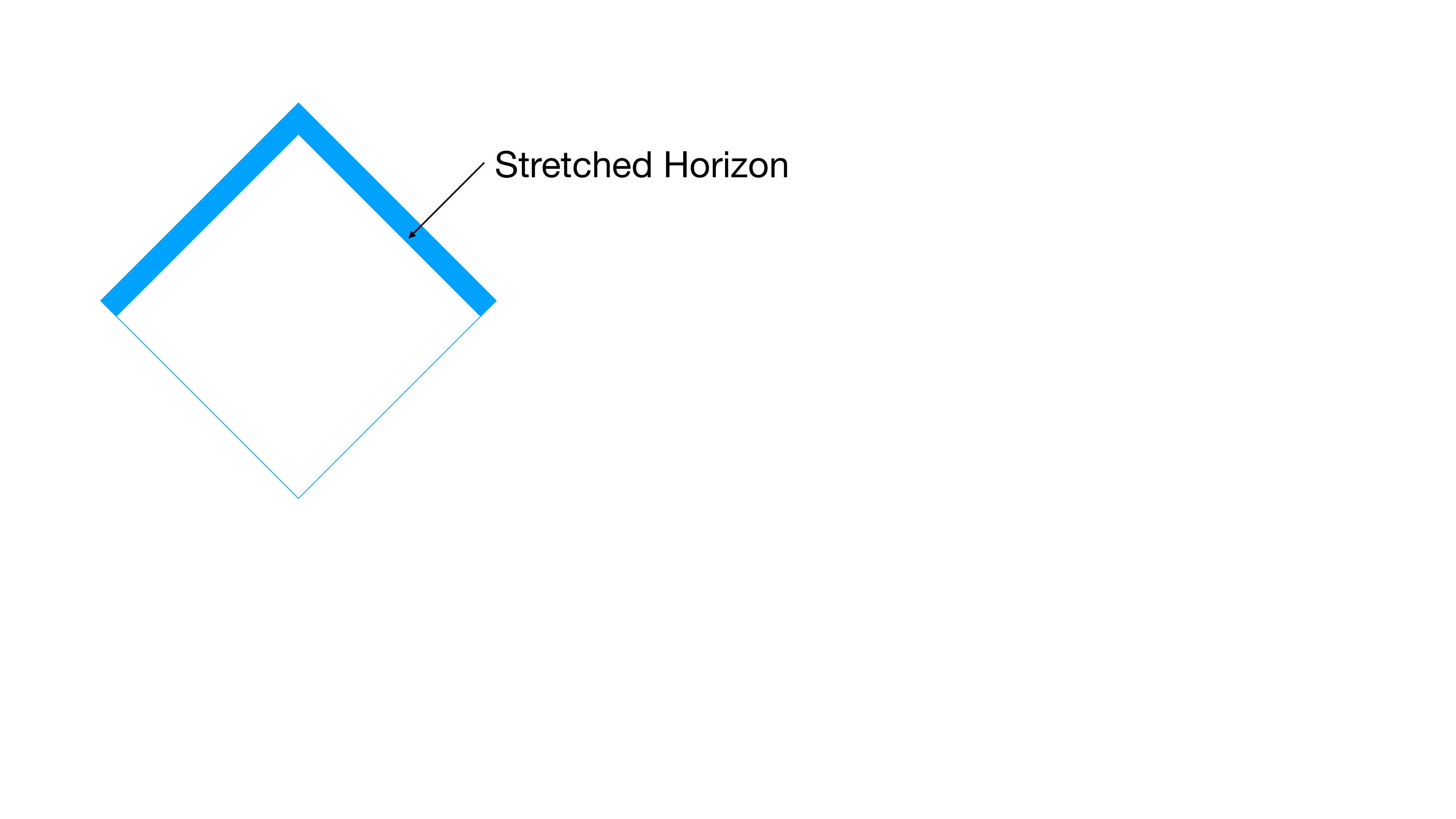}
\caption{Nested Causal Diamonds Separated by Stretched Horizon}
\label{z}
\end{center}
\end{figure}

Now consider, as in Figure 1, a pair of nested causal diamonds in a solution of the vacuum Einstein equations, whose future tips are separated by one Planck interval of proper time along the geodesic that defines them both.  The region of space-time between them is the future stretched horizon of the larger diamond, and according to the CS ansatz it contains a fluctuating entropy current $S_{vv} (z)$ with two point function
\begin{equation} \langle S_{vv} (z) S_{vv} (w) \rangle = \frac{c}{4 (z - w)^4} . \end{equation} Here $z,w$ are dimensionless null coordinates along the horizon and the $v$ direction is the future directed null vector pointing out of the diamond.  $c =  \frac{A_{\diamond}}{4 G_N}$, and the singularity at $z = w$ is cut off by a $1 + 1$ dimensional momentum cutoff on the CS CFT\footnote{In the models proposed in\cite{hilbertbundles} individual Dirac fermions are restricted to about 10-20 momentum modes.}. 
The area here is that of the larger diamond.  This distinction is important because for space-time dimension $d \geq 4$ a single step in proper time causes a large increase in area, once the area is large.  According to the holographic principle, the degrees of freedom giving rise to the fluctuations all live in the vicinity of the maximal area surface on the diamond, which implies very similar values of $z$ and $w$ will be the only contributions to the integral over the stretched horizon.  It's not clear that we can determine the order $1$ coefficient in the fluctuation relation with any precision.  

As a consequence of the large increase in entropy per Planck step, we can consider the fluctuations at different Planck separated time intervals to be independent of each other.  In other words, our system is subjected to {\it white noise}.  Thus, assuming the CEP we have a high entropy hydrodynamic system subjected to white noise with time correlations on the Planck scale.  The hydrodynamic variables of a causal diamond are encoded in the metric $h_{ab}$ of the maximum volume surface on the diamond's boundary, and the extrinsic curvature $K_{ab}$ of this surface in the full space-time.  Since the surface is extremal, $K_{ab}$ is traceless.

In the vicinity of each each extremal surface in a nest of causal diamonds, we can set up Gaussian null coordinates in which the space-time metric takes the form
\begin{equation} ds^2 = - 2 du dv + h_{ab} dx^a dx^b . \end{equation}  $u$ is a past pointing null coordinate and $v$ a future pointing null coordinate.  These coordinates for a given diamond will be valid for some small region around any given diamond's extremal surface.  They can break down near caustics, or other coordinate or real singularities.  However, we can patch them together to define a smooth future pointing null coordinate $v$ in many different ways.  Independently of those choices, there is a unique smooth flow of the geometrical quantities $h_{ab}$, $K_{ab}$ (or rather the scalar invariants derived from them) as a function of the proper time of the future tip of the diamond.  The resulting coordinate system is tied to a particular geodesic, with a give choice of initial proper time, and its continuation outside the causal diamond for each final proper time is entirely a matter of convenience.  The important point, as we will see, is that the resulting proper time evolution equations of $h_{ab}$ and $K_{ab}$ are restrictions of the covariant Einstein equations to these coordinates.  

The Einstein equations define a Hamiltonian flow of the canonically conjugate variables $h_{ab}$ and
$\pi^{ab} = \sqrt{h} K^{ab}$ , with Hamiltonian
\begin{equation} H = \int d^{d-2} x\ [ h^{-1/2} \pi^{ab} \pi_{ab} - h^{1/2} (R (h) + 8\pi G_N T_{vv})]. \end{equation} $R(h)$ is the Ricci scalar of $h$.  This system is subject to the constraint 
\begin{equation} h^{ab} K_{ab} = 0 . \end{equation}  Note that this does not mean that the $ d - 2$ volume of the maximal area surface is constant along the flow.  That volume is given by
\begin{equation} A = \int d^{d - 2} x\ \sqrt{h} , \end{equation} and the coordinate position of the maximum volume surface changes with proper time along the geodesic in the coordinate systems we have chosen.   

We now want to consider the sort of calculations that have been done in the literature using Euclidean gravitational path integrals and understand how they emerge in the framework we've outlined here.  Calculations of entropies of causal diamonds are of course put in as an axiom in our approach.  The most interesting calculations are those of spectral form factors.   In general, the dynamics we have outlined has a time dependent Hamiltonian, so there is no energy spectrum to explore.  Nonetheless we can study the quantity  $|Z(t + i \beta)|^2$ where
\begin{equation} Z(t + i \beta) \equiv \langle E | U(t + i \beta) |E \rangle . \end{equation} Here $U(t)$ are the sequence of unitary embedding operators of one diamond Hilbert space into the next along a geodesic, and $ | E \rangle $ is the {\it empty diamond} state for the given space-time.  This is a hypothetical state on the full Hilbert space with the properties
\begin{equation} \langle E | K_{\diamond} | E \rangle  = (\Delta K_{\diamond})^2 = \frac{A}{4G_N} . \end{equation} for every finite area causal diamond in the space-time.  

As usual, the exact spectral form factor exactly factorizes, but in the hydrodynamic approximation, which is valid only for time averaged quantities, it consists of a disconnected and a connected part.  The disconnected part is given by the analytic continuation of the classical action for the transverse geometry.  A one loop determinant correction can also be calculated from this action.   The connected part comes entirely from the fluctuations.  The CS ansatz guarantees that these are much smaller than the classical action, so the connected contribution is suppressed by a factor of order $e^{- S_{cl}}$.  Note that in the case of a diamond whose area grows with time, this term becomes less and less important in the asymptotic future.  

Standard derivations of hydrodynamics from quantum mechanics\cite{hydroderivations} start from a formula for expectation values in a density matrix.  The hydrodynamic fields emerge as fluctuating statistical fields, satisfying diffusion, Navier-Stokes, {\it etc.} equations with fluctuating sources.  In real applications there may be non-quantum sources of fluctuation coming from perturbations of the system from external forces that don't appear in the fundamental quantum Hamiltonian. From the point of view of real time path integrals these hydrodynamic path integrals are computing approximations to path integrals on a Schwinger-Keldysh or Closed Time Path contour.  On the other hand when one is computing the spectral form factor for a time independent Hamiltonian, it's clear that the contour in imaginary time is two periodic circles, with period $\beta$.  Swingle and Winer\cite{swingwin} have argued that the same hydrodynamic action should be used on this novel contour, and that in the context of general chaotic quantum systems hydrodynamics reproduces the entire linear ramp in the SFF, up to the Heisenberg time.

The hydrodynamic action of our theory of maximal area surfaces along a given geodesic is
\begin{equation} S = \int dt\ d^{d-2} x [\pi^{ab} \partial_t h_{ab}  + \sqrt{h} (R^{(d-2)}[h] + S_{vv}) ] . \end{equation}  In order to evaluate this action for a given geodesic in a given space-time we have to find the coordinate locus in local Gaussian null coordinates of the maximal area surface for each proper time.  $h_{ab}$ is the metric on that surface and $\pi^{ab}$ is determined by the Lagrange equations of this action, if the entropy current were a fixed source.  According to the CS ansatz, the entropy current is a fluctuating source with zero mean (if the background is a solution of the vacuum Einstein equations) and a two point function determined in terms of the instantaneous area.  The sources are localized on the stretched horizon.

Following Verlinde\cite{herman} we can write the first term in the action as 
\begin{equation} \int d^2 z\ d^{(d - 2)}\Omega\ d\pi^{ab} \wedge dh_{ab}  , \end{equation} where the two dimensional Riemannian manifold with coordinates $z$ has a boundary equal to the two circles of the SFF contour.  The geometry of this manifold does not need to be specified. As first appreciated by\cite{colemanetal}\cite{sss}\cite{SSS1} a Euclidean wormhole contribution to the path integral can sometimes be thought of as a Gaussian integral over random couplings.

The question we have to answer is whether the same is true for the random stress tensor coupling $S_{vv} = 8\pi G T_{vv}$.  The first relevant remark is that despite the apparently local nature of this coupling, the Carlip-Solodukhin derivation shows that the stress tensor two point function is completely non-local on the transverse dimensions.  The stress tensor correlators are those of a scalar field living on the stretched horizon, with an improvement term that measures the entire area of the screen.  This non-locality is also consistent with the fast scrambling\cite{lshpss} of information on the holographic screen.  Note that the two dimensional space that we introduced to rewrite the $\int \pi \partial_t h$ term in the Lagrangian only had a topology.  We did not have to introduce any metrical structure to write it.  The Hamiltonian part of the action lives only on the boundary of the two dimensional space.  So it's really only the wormhole term that requires us to introduce a two dimensional metric and an action for it.  

Integrating over Gaussian fluctuations of the $S_{vv}$, assuming the CS ansatz, we get a term in the effective action of the form 
\begin{equation} \delta S = \int d^2 t\ d^{d-2} x\ d^{d-2} y\  \alpha \frac{A(t)}{4G_N} \sqrt{h(t,x^a)}\sqrt{h(-t,y^a)} . \end{equation}  The dimensionless parameter $\alpha$ is determined by the cutoff of the singularity the two dimensional CFT two point function in the CS ansatz.  In\cite{hilbertbundles} this was conjectured to be related to the size of the causal diamond for which the CS ansatz breaks down.  All indications are that in known models of quantum gravity this is highly model dependent because it depends on the values of higher order curvature corrections to the Einstein Hilbert action.  

 Like traditional Euclidean wormhole contributions, our fluctuation action is non-local in the transverse directions. It is however, localized in the two light cone directions.  The longitudinal localization (the $u$ direction) is a consequence of our assumption that the fluctuations are restricted to the Planck scale {\it stretched horizon} region.   Localization in $v$ is localization in proper time along the geodesic and this means that our equation is analogous to the collective field equations for the {\it Brownian SYK Model}\cite{SSS1}.  A major difference from Brownian SYK is that there is only one fluctuating operator in the effective theory: the total volume of the transverse geometry. 

The other major difference between our generic story and models of black holes or fixed quantum chaotic systems is that we are talking about a history described by unitary embeddings, rather than unitary evolution on a fixed Hilbert space.  More properly\cite{hilbertbundles} we're discussing a part of the unitary evolution on one fiber of a Hilbert bundle that describes a global space-time.  The unitary embeddings describe the parts of the Hilbert space accessible to a detector on a fixed time-like trajectory. 
It is not clear whether the non-local statistical path integral we've discovered has a nice description as an integral over fields on a smooth Euclidean manifold or even something like the double cone wormhole of\cite{SSS1}\cite{maldaetal}.   What is clear is that we do not expect it to exhibit a ramp.  The number of q-bits in the system increases indefinitely with time (unless the c.c. is positive, but see remarks below about de Sitter space).  

We've obtained a statistical functional integral for a system defined in terms of nested causal diamonds of growing size along a particular time-like geodesic.  This is best thought of in terms of the Hilbert bundle formulation of quantum gravity\cite{hilbertbundles}, in which quantum mechanics must be done over all the different geodesics, with consistency conditions on the density matrices of overlapping causal diamonds.  These provide the missing information, unavailable at finite times, to the causal evolution equations.  In the hydrodynamic approximation in which we are working, the solution of the consistency conditions is obvious: it is the gravitational path integral, with any globally consistent gauge fixing.   The usual arguments about general covariance and gauge fixing tell us that we can use the covariant path integral and the BRST formalism to translate information from the Gaussian null coordinates around one set of nested causal diamonds, to any other one.  In particular, for a space that is asymptotically AdS, we can translate the fluctuating hydrodynamics of one detector to a set of statements about CFT correlators on the boundary, as long as the hydrodynamic approximation is valid for the process under consideration.  

Now consider a background configuration which is an eternal black hole.  View it first in terms of the nested causal diamonds at a fixed Schwarzschild coordinate, very close to the horizon.  Then the causal diamonds all have entropy about equal to the black hole entropy. The authors of\cite{SSS1}\cite{maldaetal} have argued that the leading fluctuation corrections to the spectral form factor, which give rise to the leading piece of its connected part, are given by the double cone Lorentzian wormhole geometry
\begin{equation}   -(\rho - i\epsilon)^2 dt^2 + d\rho^2 + d\Sigma_{d-2}^2 .\end{equation}
The singularity in this geometry is regularized by complexifying it slightly.  In the case of JT gravity the double trumpet geometry\cite{sss} is a different real slice through the same complex geometry. They both give rise to the same formula for the ramp.  We can write a path integral over the double cone geometry as a Euclidean path integral. The linear rise of the ramp comes from integrating over the collective coordinate of the relative time between the two boundaries on the double cone/double trumpet.  These formulas are of precisely the type expected from our general discussion.  In this case, since $A(t)$ is time independent, it's clear how to model the non-local action generated by integrating out the boundary entropy fluctuations by a geometric wormhole.  However, as emphasized in\cite{maldaetal} the double cone Lorentzian wormhole gives rise in higher dimensions to complex quasi-normal mode frequencies, and is not equivalent to a Euclidean wormhole.  So in general, the leading connected parts of black hole spectral form factors, are given by non-trivial wormhole configurations, but the wormholes are Lorentzian.  

So, for the case of black holes in asymptotically flat or AdS geometries we can view the gravitational path integral as computing the spectral form factor as measured by a detector at any fixed Schwarzschild position.  In asymptotically AdS space this can be continued to space-like infinity and encoded in gauge invariant boundary correlators.  In asymptotically flat space, we know that the black hole is not a stable equilibrium system. 
The hydrodynamic equations allow us to define finite time quantities in a gauge invariant manner, in the sense that gauge choices are tied to measurements made by idealized detectors, which detect only coarse grained properties.  Such detectors are not as difficult to model as systems that pretend to measure maximal amounts of quantum information.  Thus, we can discuss the finite time spectral form factor of an unstable black hole.  In principle one could attempt to study it during the process of Hawking decay, though we will not attempt to do so in this paper.  

\subsection{Higher Topology}

Our methods cannot justify the inclusion of higher genus topologies in the sum except by the indirect argument that in the case of black holes or other geometries with time-like Killing vectors they could be interpreted as the first term in the expansion of a random matrix ensemble, with a known spectral density.  For unstable black holes, the higher genus terms are of little interest because they become important on time scales long compared to the black hole evaporation time.  Only a part of the ramp is, in principle, observable.  The same is true for a hypothetical stable finite entropy de Sitter space, because no localized detector stays inside the causal diamond of any given de Sitter geodesic for longer than a de Sitter scrambling time.  The internal clocks of long lived detectors experience time dependent physics idiosyncratic to the particular local group of galaxies in which the detector resides.  It is inconsistent to treat the physics measured by such a detector by the methods appropriate to random time independent Hamiltonians.

Thus higher genus topological contributions are relevant, at best, to black holes in AdS/CFT models with Einstein-Hilbert duals, and only if the random matrix assumption is valid.  One should also be aware of the fact that corrections to the density of states from higher terms derivative expansion of the action, might be more important than the higher genus corrections with the leading order density of states.

\section{Conclusions}

We have given a purely hydrodynamic derivation of gravitational path integrals including some with non-trivial topology, starting from three postulates about the underlying hydrodynamic variables
\begin{itemize}
\item  A solution of the vacuum Einstein equations represents the equilibrium hydrodynamic state of a quantum system whose subsystems are represented by causally disjoint causal diamonds.  The modular Hamiltonian of any causal diamond satisfies\cite{ted95}\cite{carlip}\cite{solo}\cite{BZ}
\begin{equation} \langle K \rangle = (\Delta K)^2 = \frac{A}{4G_N}, \end{equation} where $A$ is the maximal $d -2$ volume on the diamond boundary.  
\item The hydrodynamic variables are $h_{ab}$, the geometry on this maximal surface, and its canonical conjugate $\pi^{ab}$ with respect to proper time flow along a geodesic sweeping out a family of nested causal diamonds in the space-time. $\pi^{ab}$ is proportional to the extrinsic curvature of the surface. The flow is Hamiltonian, with a Hamiltonian derived from Einstein's equations in coordinates that reduce to Gaussian null coordinates in the neighborhood of each maximal surface.  In addition, there is a coupling 
\begin{equation} \delta H = \int d^{d-2} x \sqrt{h(x^a,t)} S_{vv} (x^a,t) , \end{equation} to the fluctuating entropy current on the boundary of the diamond.  The two point function of $S_{vv}$ is independent of $x^a$  and proportional to $\frac{A(t)}{4G_N}$ with a coefficient that would be singular in the absence of a cutoff on the CS CFT.  Thus, it looks like a non-local ``wormhole" contribution to the {\it statistical} gravitational path integral over $h_{ab}$, in the transverse directions.  It is however, localized in the two light cone directions.  The longitudinal localization (the $u$ direction) is a consequence of our assumption that the fluctuations are restricted to the Planck scale {\it stretched horizon} region.   Localization in $v$ is localization in proper time along the geodesic and this means that our equation is analogous to the collective field equations for the {\it Brownian SYK Model}\cite{SSS1}.  A major difference from Brownian SYK is that there is only one fluctuating operator in the effective theory: the total volume of the transverse geometry.  
\item For black holes and other geometries with a Killing symmetry and a fixed area horizon, the fluctuation term can be modeled by a two dimensional double cone Lorentzian wormhole, following the work of\cite{SSS1}\cite{maldaetal}.  Using a trick of\cite{herman} we can also write the kinetic term of the fluctuating metric $h_{ab}$ as an integral over the two dimensional space.  
\item The consistency of the description of the black hole along any geodesic in the background space-time is ensured by noting that the fluctuating equations are a gauge fixed version of a gravitational path integral, which is manifestly generally covariant.  
\end{itemize}

The purpose of this note has been to show how, in some simple cases, gravitational path integral formulae for connected two point correlators of the complexified time evolution operator can be derived from a few statistical assumptions connecting the properties of quantum density matrices of causal diamonds in a background space-time to the maximal volume on the boundary of the diamond.  These assumptions are {\it background dependent}, and {\it hydrodynamic in nature}.  They give no justification for using the GPI as a microscopic quantum mechanical definition of quantum gravity.   In light of these results, we should view the use of the GPI to compute graviton amplitudes in asymptotically flat and AdS backgrounds, as analogs of the use of the Navier-Stokes equations to define perturbative phonon interactions in condensed matter physics.  

The results also give no justification for the idea that quantum gravity requires a sum over all topologies of {\it .e.g. } Euclidean manifolds. This is a good thing, since such a sum is ill defined in any space-time dimension above $3$.  Recent beautiful results relating sums over three dimensional topologies to averages over $2$ dimensional CFTs have been explained by Collier and collaborators as calculating a universal sector of the partition function of high dimensional operators in CFTs with large central charge\footnote{S. Collier, private communication.} .   Instead, one sums only over two dimensional topologies and the extension from general Hamiltonian systems with fluctuations to two dimensional wormhole topologies has been shown in an elegant manner by Verlinde\cite{herman}.  The question of whether and how to include higher genus two dimensional topologies depends on the extent to which one believes that the time dependent behavior of the gravitational system in question obeys random matrix statistics with fixed entropy.  We've argued that there are few realistic gravitational examples where this is the case, but it could well be true for stable black holes in AdS space.  To analyze those in higher dimension one would have to understand how to generalize the Lorentzian double cone to higher genus.  

It's clear that black holes in Minkowski or de Sitter space decay long before anything but the linear ramp behavior could become evident.  de Sitter space, if it is indeed stable, is a finite entropy system.  However, there is no detector that measures the time along a geodesic for a time longer than $\sim R_{dS} {\rm ln}\ \frac{R_{dS}}{L_P}$\footnote{In addition, no detector can, even in principle, measure most of the q-bits of the dS Hilbert space.}.  Thus, mathematical predictions about dS recurrence times are totally void of scientific meaning\cite{tbwfsp}, and again only the ramp has any chance of being a true physical observable.

Thus, the predictions based on higher genus topologies are useful predictions for field theory systems that have some kind of AdS dual.  These predictions might in principle be verified by laboratory experiments on condensed matter systems.   They have no relevance to quantum gravity applications in the real world.  

 \section{Appendix: Why is AdS Different?}
 
 Our emphasis in this note has been on non-negative c.c. or causal diamonds smaller than the AdS radius in models with negative c.c. .  In order to understand why negative c.c. is globally so different, we have to recall the role of the c.c. as the regulator of the asymptotics of the two measures of the size of causal diamonds: proper time and holographic screen area.  The crucial point is that for negative c.c. the area goes to infinity at finite proper time along any time-like geodesic.  After that point there is an infinite redshift factor between proper time and time on the conformal boundary.  
 
 In asymptotically flat space the area goes to infinity in conjunction with the proper time, and the quantum number defined by entropy deficits on causal diamond boundaries becomes an asymptotically conserved Hamiltonian, but there is no further evolution on the conformal boundary.  The AdS/CFT correspondence has taught us that the infinities of the negative c.c. case are handled very differently.  This is extremely transparent in the tensor network formulation, where the limit of infinite entropy corresponds to taking the lattice spacing to zero in a lattice approximation to a CFT.  When we do this, most of the states of the lattice model are thrown away, and the high entropy density matrix one might have tried to assign to a flat space region of comparable size is not the correct `` empty diamond state".   Instead, according to the TNRG, we've been coupling together nodes with nearest neighbor couplings and finding a pure lattice state that approximates the CFT ground state.   
 
 Thus for non-negative c.c. the empty diamond state of the global geometry is impure and saturates the covariant entropy bound\cite{fsb}.  States with any localized excitation in any diamond are low entropy constrained states, and the energy is a count of the number of constrained q-bits, which becomes asymptotically conserved in flat space because the number of constrained q-bits goes to infinity for large diamonds.  For negative c.c. the empty diamond state is pure, but when looked at on a sufficiently microscopic scale the local diamonds behave like diamonds for non-negative c.c..  The model that captures this structure is a lattice approximation to the CFT in which each lattice point is thought of as the bifurcation surface of a causal diamond of radius slightly smaller than $R_{AdS}$.  This sort of model nicely explains the different properties of stable AdS black holes and those for non-negative c.c. or of sub-AdS radius size.  The lattice structure explains the existence of sound modes in the hydrodynamic spectrum and the different coefficient in the relation between entanglement capacity and entropy.   It also explains why the infall time to the singularity is always the AdS radius, rather than the black hole size, if we interpret the singularity as thermalization and imagine that the nodes of the lattice are fast scramblers.  
 

\begin{thebibliography}{99}
 \bibitem{sss} P.~Saad, S.~H.~Shenker and D.~Stanford, ``JT gravity as a matrix integral,''
[arXiv:1903.11115 [hep-th]].
 \bibitem{Feynman} Feynman, R. P.; Hibbs, A. R.; Styer, D. F. (2010). Quantum Mechanics and Path Integrals. Mineola, NY: Dover Publications. pp. 29Ð31. ISBN 978-0-486-47722-0.
 \bibitem{KacWiener}  Kac, Mark (1949). "On Distributions of Certain Wiener Functionals". Transactions of the American Mathematical Society. 65 (1): 1Ð13. doi:10.2307/1990512. 
 \bibitem{tbwormhole}T.~Banks,``Fluctuating Hydrodynamics and Wormholes,''
[arXiv:2203.08855 [hep-th]].
 \bibitem{ted95} T.~Jacobson,
``Thermodynamics of space-time: The Einstein equation of state,''
Phys. Rev. Lett. \textbf{75}, 1260-1263 (1995)
doi:10.1103/PhysRevLett.75.1260
[arXiv:gr-qc/9504004 [gr-qc]].
\bibitem{carlip} S.~Carlip, ``Black hole entropy from conformal field theory in any dimension,''
Phys. Rev. Lett. \textbf{82}, 2828-2831 (1999)
doi:10.1103/PhysRevLett.82.2828
[arXiv:hep-th/9812013 [hep-th]];
\bibitem{solo}S.~N.~Solodukhin,
``Conformal description of horizon's states,''
Phys. Lett. B \textbf{454}, 213-222 (1999)
doi:10.1016/S0370-2693(99)00398-6
[arXiv:hep-th/9812056 [hep-th]].
\bibitem{BZ} T.~Banks and K.~M.~Zurek,
``Conformal description of near-horizon vacuum states,''
Phys. Rev. D \textbf{104}, no.12, 126026 (2021)
doi:10.1103/PhysRevD.104.126026
[arXiv:2108.04806 [hep-th]].
\bibitem{tedentangle} T.~Jacobson,
``Entanglement Equilibrium and the Einstein Equation,''
Phys. Rev. Lett. \textbf{116}, no.20, 201101 (2016)
doi:10.1103/PhysRevLett.116.201101
[arXiv:1505.04753 [gr-qc]].
\bibitem{fsb} W.~Fischler and L.~Susskind, ``Holography and cosmology,''
[arXiv:hep-th/9806039 [hep-th]];
R.~Bousso,``A Covariant entropy conjecture,''
JHEP \textbf{07}, 004 (1999)
doi:10.1088/1126-6708/1999/07/004
[arXiv:hep-th/9905177 [hep-th]];
R.~Bousso,
``Holography in general space-times,''
JHEP \textbf{06}, 028 (1999)
doi:10.1088/1126-6708/1999/06/028
[arXiv:hep-th/9906022 [hep-th]];
R.~Bousso,
``The Holographic principle for general backgrounds,''
Class. Quant. Grav. \textbf{17}, 997-1005 (2000)
doi:10.1088/0264-9381/17/5/309
[arXiv:hep-th/9911002 [hep-th]].
\bibitem{sorkin} R.~D.~Sorkin,
``1983 paper on entanglement entropy: ''On the Entropy of the Vacuum outside a Horizon'',''
[arXiv:1402.3589 [gr-qc]].
\bibitem{srednickietal} M.~Srednicki, ``Entropy and area,''
Phys. Rev. Lett. \textbf{71}, 666-669 (1993)
doi:10.1103/PhysRevLett.71.666
[arXiv:hep-th/9303048 [hep-th]];
C.~G.~Callan, Jr. and F.~Wilczek,``On geometric entropy,''
Phys. Lett. B \textbf{333}, 55-61 (1994)
doi:10.1016/0370-2693(94)91007-3
[arXiv:hep-th/9401072 [hep-th]];
L.~Susskind and J.~Uglum,
``Black hole entropy in canonical quantum gravity and superstring theory,''
Phys. Rev. D \textbf{50}, 2700-2711 (1994)
doi:10.1103/PhysRevD.50.2700
[arXiv:hep-th/9401070 [hep-th]];
T.~Jacobson,
``Black hole entropy and induced gravity,''
[arXiv:gr-qc/9404039 [gr-qc]].

\bibitem{bfm}T.~Banks, B.~Fiol and A.~Morisse,
``Towards a quantum theory of de Sitter space,''
JHEP \textbf{12}, 004 (2006)
doi:10.1088/1126-6708/2006/12/004
[arXiv:hep-th/0609062 [hep-th]].
\bibitem{hst} T.~Banks and W.~Fischler,
``Holographic Theory of Accelerated Observers, the S-matrix, and the Emergence of Effective Field Theory,''
[arXiv:1301.5924 [hep-th]];
T.~Banks and W.~Fischler,
``No Firewalls in Holographic Space-Time or Matrix Theory,''
[arXiv:1305.3923 [hep-th]];
T.~Banks and W.~Fischler,
``Holographic Space-time and Newton's Law,''
[arXiv:1310.6052 [hep-th]];
T.~Banks and W.~Fischler,
``Holographic Space-time, Newton's Law and the Dynamics of Black Holes,''
[arXiv:1606.01267 [hep-th]];
T.~Banks and W.~Fischler,
``Holographic space-time, Newton{\textquoteright}s law, and the dynamics of horizons,''
Adv. Theor. Math. Phys. \textbf{27}, no.1, 65-86 (2023)
doi:10.4310/ATMP.2023.v27.n1.a3
[arXiv:2003.03637 [hep-th]].
\bibitem{hydroderivations} See the many references to earlier work in T.~Banks and A.~Lucas, ``Emergent entropy production and hydrodynamics in quantum many-body systems,''
Phys. Rev. E \textbf{99}, no.2, 022105 (2019)
doi:10.1103/PhysRevE.99.022105
[arXiv:1810.11024 [cond-mat.stat-mech]], where hydrodynamic equations for a large family of quantum systems on graphs.
\bibitem{swingwin} M.~Winer and B.~Swingle,
``Hydrodynamic Theory of the Connected Spectral form Factor,''
Phys. Rev. X \textbf{12}, no.2, 021009 (2022)
doi:10.1103/PhysRevX.12.021009
[arXiv:2012.01436 [cond-mat.stat-mech]].
\bibitem{herman} H.~Verlinde,``Wormholes in Quantum Mechanics,''
[arXiv:2105.02129 [hep-th]].
\bibitem{colemanetal} S.~R.~Coleman,``Black holes as red herrings: Topological fluctuations and the loss of quantum coherence,''
Nucl. Phys. B \textbf{307}, 867-882 (1988)
doi:10.1016/0550-3213(88)90110-1
T.~Banks,
``Prolegomena to a Theory of Bifurcating Universes: A Nonlocal Solution to the Cosmological Constant Problem Or Little Lambda Goes Back to the Future,''
Nucl. Phys. B \textbf{309}, 493-512 (1988)
doi:10.1016/0550-3213(88)90455-5
S.~B.~Giddings and A.~Strominger,
``Baby Universes, Third Quantization and the Cosmological Constant,''
Nucl. Phys. B \textbf{321}, 481-508 (1989)
doi:10.1016/0550-3213(89)90353-2
\bibitem{SSS1} P.~Saad, S.~H.~Shenker and D.~Stanford,
``A semiclassical ramp in SYK and in gravity,''
[arXiv:1806.06840 [hep-th]].
\bibitem{maldaetal} Y.~Chen, V.~Ivo and J.~Maldacena,
``Comments on the double cone wormhole,''
JHEP \textbf{04}, 124 (2024)
doi:10.1007/JHEP04(2024)124
[arXiv:2310.11617 [hep-th]].
\bibitem{lshpss} L.~Susskind and J.~Lindesay,
``An introduction to black holes, information and the string theory revolution: The holographic universe,''
P.~Hayden and J.~Preskill, ``Black holes as mirrors: Quantum information in random subsystems,''
JHEP \textbf{09}, 120 (2007)
doi:10.1088/1126-6708/2007/09/120
[arXiv:0708.4025 [hep-th]];
Y.~Sekino and L.~Susskind, ``Fast Scramblers,''
JHEP \textbf{10}, 065 (2008)
doi:10.1088/1126-6708/2008/10/065
[arXiv:0808.2096 [hep-th]].
\bibitem{hilbertbundles} T.~Banks,
``Hilbert Bundles and Holographic Space-time: the Hydrodynamic Approach to Gravity,''
[arXiv:2502.04924 [hep-th]];
T.~Banks,
``Hilbert Bundles and Holographic Space{\textendash}Time Models,''
Astronomy \textbf{4}, no.2, 7 (2025)
doi:10.3390/astronomy4020007
[arXiv:2306.07038 [hep-th]].
\bibitem{tbwfsp} T.~Banks, W.~Fischler and S.~Paban,
``Recurrent nightmares? Measurement theory in de Sitter space,''
JHEP \textbf{12}, 062 (2002)
doi:10.1088/1126-6708/2002/12/062
[arXiv:hep-th/0210160 [hep-th]].
\end{thebibliography}
\end{document}